\documentclass[aps,prl,twocolumn,showpacs,superscriptaddress]{revtex4}
\usepackage{graphicx}
\usepackage{bm}

\newcommand{\beq}{\begin{equation}}
\newcommand{\eeq}{\end{equation}}

\begin{document}

\title{\boldmath
Phase Transitions in Nucleonic Matter and Neutron-Star Cooling
}

\author{V.~A.~Khodel}
\affiliation{
Russian Research Centre Kurchatov Institute,
Moscow, 123182, Russia
}
\affiliation{
McDonnell Center for the Space Sciences and
Department of Physics, Washington University,
St.~Louis, MO 63130, USA
}

\author{J.~W.~Clark}
\affiliation{
McDonnell Center for the Space Sciences and
Department of Physics, Washington University,
St.~Louis, MO 63130, USA
}

\author{M.~Takano}
\affiliation{
Advanced Research Institute for
Science and Engineering, Waseda University,
3-4-1 Okubo Shinjuku-ku, Tokyo 169-8555, Japan
}
\author{M.~V.~Zverev}
\affiliation{
Russian Research Centre Kurchatov Institute,
Moscow, 123182, Russia
}

\date{\today}

\begin{abstract}
A new scenario for neutron-star cooling is proposed, based on the
correspondence between pion condensation, occurring in neutron matter
due to critical spin-isospin fluctuations, and the metal-insulator
phase transition in a two-dimensional electron gas.  Beyond the
threshold density for pion condensation, where neutron-star matter
loses its spatial homogeneity, the neutron single-particle spectrum
acquires an insulating gap that quenches neutron contributions
to neutrino-production reactions and to the star's specific
heat.  In the liquid phase at densities below the transition point,
spin-isospin fluctuations are found to play dual roles.  On the 
one hand, they lead to a multi-sheeted neutron Fermi surface that 
extends to low momenta, thereby activating the normally forbidden 
direct-Urca cooling mechanism; on the other, they amplify the 
nodeless $P$-wave neutron superfluid gap while suppressing $S$-wave 
pairing.  In this picture, lighter stars without a pion-condensed 
core experience slow cooling, while enhanced cooling occurs in heavier 
stars through direct-Urca emission from a narrow shell of the interior.

\end{abstract}

\maketitle

Observations designed to detect and measure thermal radiation from 
neutron stars with ages between $10^2$ and $10^5$ years can provide 
valuable information on the internal temperatures $T$ of these stars 
at the stage when cooling proceeds via neutrino emission from their 
neutrino-transparent cores \cite{yak0}.  Such information may be 
used to constrain theories of hadronic matter at high density.  The 
relevant neutrino reactions are (i) the so-called direct Urca (DU) 
process, in which beta decay and inverse beta decay operate in tandem
on thermally activated nucleons and electrons, (ii) the modified 
Urca (MU) process, involving a nucleon spectator in the Urca reactions, 
(iii) neutrino bremsstrahlung from nucleon-nucleon collisions, and 
(iv) neutrino emission due to Cooper pairing in the superfluid phase 
\cite{flowers,yak1}.  Of these, only the DU mechanism can produce 
rapid cooling, as the data seems to require for some stars.  However, 
its operation is normally forbidden by the large mismatch of neutron 
and proton Fermi momenta, which can be compensated by spectators 
in the MU process generally associated with slow cooling.  The 
importance of the Cooper-pairing mechanism (iv) is highly sensitive 
to the proton and neutron gaps, whose values in turn depend crucially 
on the internal composition and structure of the star.

A new perspective on neutron-star interiors is suggested by
laboratory studies of (i) the metal-insulator transition in
two-dimensional (2D) silicon samples of low disorder \cite{pud},
and (ii) the liquid-solid phase transition in 2D $^3$He 
\cite{godfrin}.  Quite far from the transition point, 
the electron and $^3$He systems involved may be described as 
normal Fermi liquids obeying Landau theory.  Beyond the
transition, these systems become inhomogeneous, with an insulating gap
in the single-particle (sp) spectrum.  Similar behavior of the
sp spectra has been observed in the ``normal'' states of
high-$T_c$ superconductors \cite{shen}.
Cold neutron matter also behaves as a normal (or superfluid) Fermi 
liquid at relatively low densities.  With increasing depth in the star 
and coincident increase in baryon density $\rho$, spin-isospin 
fluctuations grow in importance.  At a critical density $\rho_c$, 
a condensate of spin-isospin excitations, specified by critical wave 
number $q_c\sim p_{Fn}$, forms in the channel with $\pi^0$ quantum 
numbers \cite{mig,dyug}.  State-of-the-art
microscopic calculations \cite{wir} predict that $\pi^0$
condensation (PC) sets in at a density comparable to the equilibrium
density $\rho_0=0.16$ fm$^{-3}$ of symmetrical nuclear matter,
while earlier phenomenological estimates put $\rho_c$
in the range 0.3 -- 0.4 fm$^{-3}$ \cite{mig,dyug} .
Comparing these values to the central density $\rho(0)\simeq (0.5-1.0)$
fm$^{-3}$ \cite{wir} of a typical neutron star, we infer that
at low $T$, a significant part of the stellar bulk exists in
the inhomogeneous PC phase, rather than as a Fermi liquid.

Accordingly, in contrast to the pionic cooling scenario reviewed in
Ref.~\cite{tsuruta}, we propose that in the domain occupied by the
PC, the neutron spectrum acquires an insulating gap exceeding
in value any expected superfluid gaps in neutron matter.  It
follows that in this region of the star all cooling processes requiring
the participation of neutrons are strongly inhibited.  The neutron
contribution to the specific heat $C(T)$ is suppressed as well,
since its leading term $\sim T^3$ now derives from the phonon
spectrum rather than from the gapped sp excitations.  The properties
of the embedded proton subsystem are not affected so dramatically.
Analogously to the electronic subsystem in solids, the protons
form a ``conductivity band'' with a sp spectrum of unaltered shape,
specified by an effective mass, and the behavior of their contribution
$C_p(T)\sim T$ to the specific heat remains unchanged.  Thus, in the
new cooling scenario, the large region of the star in which the
$\pi^0$ condensate holds sway becomes irrelevant to the
cooling process, except for the effects of neutrino-generating
reactions involving protons as the only nucleonic participants,
along with the proton contribution to the stellar specific heat.

As the internal temperature drops, the proton
subsystem undergoes a superconducting phase
transition, which we assume to originate from phonon-induced
attraction.  A rough estimate of the critical temperature,
$T_c^p\simeq 20$ keV, is then obtained in terms of the proton 
Fermi energy in the same way as for ordinary superconductors.

Having addressed the organization of neutron-star matter at
moderately high densities, let us now analyze what is transpiring
in the domain $0.5\,\rho_0<\rho<\rho_c$ corresponding to the outer core,
where the stellar material is a neutron liquid with fluid admixtures
of protons and neutralizing leptons.  We argue that in this region,
spin-isospin fluctuations have a strong impact on neutron pairing,
suppressing it in the $S$-wave channel and enhancing it in $P$-waves.
This situation is familiar from the physics of superfluid $^3$He, where
spin fluctuations play the key role in promoting $P$-pairing over
$S$-pairing \cite{voll}.  To estimate the neutron triplet-$P$ gap,
we adopt the BCS formalism and write the gap equation as
\beq
{\hat\Delta}({\bf p})=
-\int \left[
{\cal V}({\bf p},{\bf p}'){+}
{\cal V}^{\pi}\left({\bf p}{-}{\bf p}';\omega=E({\bf p'})\right)
\right]
{{\hat\Delta}({\bf p}')\over 2E({\bf p}')} d\tau'  \,  ,
\label{bcs01}
\eeq
with $d\tau=d^3p/(2\pi)^3$.  In the triplet $P$-wave channel, the gap
function has the form
${\hat\Delta}({\bf p})=id_{ik}p_k\sigma_i\sigma_2$, with coefficients
$d_{ik}$ yet to be determined.  The quasiparticle energy $E({\bf p})$
is given by $E({\bf p})=\left[\xi^2(p)+{\rm Tr}({\hat \Delta}({\bf p})
{\hat\Delta}^\dagger({\bf p}))\right]^{1/2}$, where $\xi(p)$ is
the sp spectrum of the normal Fermi liquid relative to the chemical
potential $\mu$, and ${\rm Tr}({\hat \Delta}({\bf p}){\hat\Delta}^
+({\bf p}))= \Delta^2+\sum_m a_mY_{2m}({\bf n})$.  In triplet
$P$-pairing, the regular interaction ${\cal V}$ is nearly exhausted 
by the frequency- and density-independent spin-orbit component of 
the scattering amplitude.  The fluctuation contribution to the
pairing interaction is written as \cite{mig}
${\cal V}^{\pi}_{\alpha\beta,\gamma\delta}(q,\omega)=\lambda^2_n( q)
({\bm \sigma}_{\alpha\gamma}\cdot{\bf q})\,{\rm Re}\,D(q,\omega)
({\bm \sigma}_{\beta\delta}\cdot{\bf q})/m_{\pi}^2$, where $\lambda_n$
is an effective charge accounting for the renormalization of the
associated vertex part.  The propagator $D$ is given by the
Migdal formula $-D^{-1}(q,\omega)= q^2+m^2_{\pi}+
\Pi_{NI}(q,\omega=0)+\Pi_{NN}(q,\omega)$, made up of the ordinary
part $\Pi_{NN}(q,\omega)=-Bq^2-iq|\omega|M^2/(2\pi m^2_{\pi})$
of the pion polarization operator $\Pi$, along with the term
$\Pi_{NI}(q,\omega=0)=-q^2\rho/\rho_I(1+q^2/q^2_I)$ arising
from pion conversion into a $\Delta$-isobar and neutron hole.
In the domain of critical fluctuations, we have \cite{mig}
\beq
- D^{-1}(q{\to}q_c;\rho{\to}\rho_c;\omega{=}0)
= \gamma^2{(q^2{-}q^2_c)^2\over q^2_I}+ \eta\kappa^2q^2_I  \  ,
\label{denom}
\eeq
where $\eta=(\rho_c-\rho)/\rho_c$.  The parameters $\gamma$
and $\kappa$ are determined from the obvious relations
$(1-B)q^2_c+m^2_{\pi}-q^2_cr_c\zeta_c=0$ and $1-B-r_c\zeta_c^2=0$,
with $r_c=\rho_c/\rho_I$ and $\zeta_c=(1+q^2_c/q^2_I)^{-1}$.
Simple algebra leads to
$ q_c=\left(m_{\pi}q_I\right)^{1/2}(1-B)^{-1/4}$,
$r_c=\left(m_{\pi}/q_I+\sqrt{1-B}\right)^2$,
$\gamma^2=r_c\zeta_c^3$,
and $\kappa^2=r_c\zeta_cq^2_c/q^2_I$.  Employing the parameter set
$q_I^2=5\,m^2_{\pi}, \rho_I\simeq 1.8\,\rho_0$, and $B\simeq 0.7$
from Ref.~\cite{mig}, one arrives at
$q_c\simeq 1.9\,m_{\pi}$, $\rho_c\simeq 2\,\rho_0$, $\gamma\simeq 0.4$,
and $\kappa\simeq 0.7$.  As will be seen, the contribution to the
gap value from spin-orbit forces is insignificant; with its neglect
Eq.~(\ref{bcs01}) is finally recast as
\begin{eqnarray}
\sum_k d_{ik}p_k=-\int {\lambda^2_n(q) \over m^2_{\pi}}
\Biggl(\sum_k d_{ik} q^2p'_k -2\sum_{k,l} d_{kl} q_i q_k p'_l\Biggr)
\nonumber \\
 \times\,  {{\rm Re}D(q,|\xi(p')|)\over 2E({\bf p}')}\,d\tau' \ ,
\qquad\qquad\qquad
\label{set}
\end{eqnarray}
where ${\bf q}={\bf p}-{\bf p}'$.  In arriving at this result,
we have made use of the relations
$\sum_{\delta}(\sigma_l\sigma_i\sigma_2)_{\alpha\delta}
(\sigma_m)_{\beta\delta}=
(\sigma_l\sigma_i\sigma_2\sigma^+_m)_{\alpha\beta}=
-(\sigma_l\sigma_i\sigma_m\sigma_2)_{\alpha\beta}$.

\begin{figure}[t]
\includegraphics[width=0.8\linewidth]{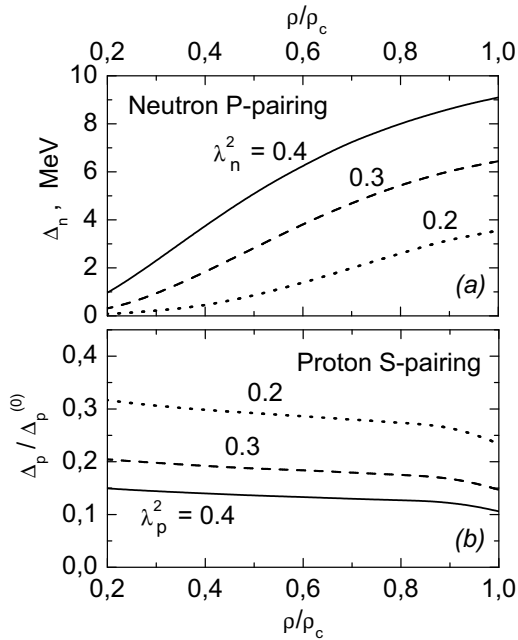}
\caption{
Neutron $P$-wave gap $\Delta_n$ at the Fermi surface (in MeV)
(panel (a)) and suppression factor $\Delta_p/\Delta^{(0)}_p$
for proton $S$-wave gap (panel (b)), versus baryon density $\rho$
in units of $\rho_c$.  (The proton fraction is taken as 0.06.)
Values of $\lambda^2_n$ and $\lambda^2_p$
are indicated by numbers near the corresponding curves.
}
\label{fig:gap}
\end{figure}

Unfortunately, the pairing problem (\ref{set}) still defies full
solution.  In neutron matter, only the spectrum of solutions
restricted to the $^3P_2$--$^3F_2$ channel has been explicated
\cite {khod1}.  The solution set contains both nodeless and nodal
combinations of the different basis states.  At $T\ll T_c$, a
nodeless solution wins the energetic competition, and we anticipate
that this feature, also inherent in superfluid $^3$He, is present
here as well.  For nodeless solutions, an adequate
approximation to the gap in the sp spectrum can be
obtained by retaining only the $\Delta^2$ term in
${\rm Tr}({\hat \Delta}({\bf p}){\hat\Delta}^+({\bf p}))$.
The matrix $d_{ik}$ then becomes proportional to
$\delta_{ik}$ and the angle integration is obviated, giving
rise to the expression $q^2\,({\bf p}\cdot{\bf p}')
- 2 ({\bf q}\cdot{\bf p})({\bf q}\cdot{\bf p}')$ in the
numerator of the r.h.s.\ of Eq.~(\ref{set}).
Furthermore, the system (3) becomes decoupled,
and we are left with a single integral equation to solve.  Exploiting the
fact that the propagator $D(q,\omega=0)$ is peaked at $q=q_c<2p_{Fn}$,
this equation is simplified to
\beq
1={ \lambda^2_n q^2_cM\over 8\pi\gamma\kappa m^2_{\pi}p_F}
\! \int\limits_0^{\infty} \! {\rm Re}\,{d\xi\over\left[\left(\eta{+}iM^2\xi/
2\pi\kappa^2 m^2_{\pi}q_c\right)
\left(\xi^2{+}\Delta^2\right)\right]^{1/2}} \ ,
\label{simbcs}
\eeq
where $\lambda^2_n \equiv \lambda^2_n(q_c)$.  In deriving this formula,
we have made the replacements
${\bf p}\cdot{\bf p}'=(p^2+(p')^2-q^2)/2 {\to} p^2_F{-}q^2_c/2$,
${\bf q}\cdot{\bf p}=p^2-{\bf p}\cdot{\bf p}' {\to} q^2_c/2$, and
${\bf q}\cdot{\bf p}'={\bf p}\cdot{\bf p}'-(p')^2 {\to} -q^2_c/2$;
their validity has been confirmed in numerical calculations.  Since 
the parameter $\lambda_n$ is as yet uncertain, numerical calculations 
of the gap $\Delta(\rho)$ have been performed for three different 
values, with the results shown in Fig.~\ref{fig:gap}. We see that 
the $P$-wave neutron gap is dramatically magnified in comparison 
with standard estimates of $S$-wave and $P$-wave gaps \cite{vladimir} 
that ignore spin-isospin fluctuations, justifying our neglect 
of the regular interaction.

Next we assess the inhibitory effect of spin-isospin fluctuations
on proton $S$-wave pairing.  To estimate the suppression factor,
we appeal to the property
${\langle}S{=}0|\sigma_i^1\sigma_k^2|S{=}0{\rangle}=-\delta_{ik}$
and multiply both sides of Eq.~(\ref{bcs01}), rewritten for the
full proton $S$-wave gap function $\Delta_p$, by $\Delta^{(0)}_p/2E_0(p)$,
where $E_0(p){=}\left[\xi^2(p)+ (\Delta^{(0)}_p)^2\right]^{1/2}$ and
$\Delta^{(0)}_p$ is the gap value found \cite{vladimir} when critical 
fluctuations are neglected.  Integration over the intermediate proton
momentum and simple manipulations utilizing the gap equation for
$\Delta^{(0)}_p$ lead to 
\beq
\int \left({1\over E_0(\xi)}-{1\over E(\xi)}\right)d\xi=
\int\!\!\int {I(\xi,\xi_1)\over E_0(\xi)E(\xi_1)}d\xi d\xi_1  \  ,
\label{dif1}
\eeq
where
\beq
I(\xi,\xi_1)= {\lambda_p^2M\over 16\pi^2 m^2_{\pi}p_F}
\int\limits_{|p-p_1|}^{p+p_1} q^3\,D(q;|\xi_1|) dq
\label{form1}
\eeq
with $p\equiv p(\xi)$ determined from the formula for the sp 
spectrum $\xi(p)$.  Results of numerical calculations of the ratio 
$\Delta/\Delta^{(0)}$ based on Eq.~(\ref{dif1}) are also 
presented in Fig.~\ref{fig:gap}.  Proton $S$-pairing 
in the liquid phase is suppressed over a wide density range.

The above results suggest that neutron $P$-pairing in the liquid
domain of the core is enhanced so strongly by spin-isospin fluctuations
that the neutrino emissivity due to neutron Cooper pairing, which 
behaves as \cite{yak1} $Q_{\rm nCp}\sim (\Delta/T)^6\exp(-2\Delta/T)$, 
is completely suppressed by the exponential factor.  At the same time, 
recent microscopic calculations \cite{lombardo} indicate that 
the $^1$S$_0$ proton gap in neutron-star matter depends crucially 
on the density $\rho$, falling off rapidly when $\rho$ exceeds 
$\rho_0$.  In view of the suppression factor from spin-isospin 
fluctuations found here and plotted in Fig.~1, we conclude that 
proton pairing is irrelevant to the neutrino-cooling stage.

We now turn to the role of the direct Urca process.
Despite their limitations, the available experimental data on 
the surface temperatures $T_s$ of neutron stars give evidence 
for the existence of slow and rapid cooling tracks.  It is
generally presumed that DU reactions are somehow involved in 
the rapid cooling process. Yet if one adopts the best
available equations of state of neutron matter \cite{wir}
derived from first principles, this highly efficient cooling
mechanism is precluded in all but the most massive
neutron stars, because of the large
difference between neutron and proton Fermi momenta.  However, a
novel route to the DU process in dense matter is opened by a
rearrangement of the neutron quasiparticle distribution $n(p)$ 
at a critical density $\rho_r<\rho_c$.  The rearrangement is 
precipitated by critical spin-isospin fluctuations seething 
in the neutron liquid near its inner boundary with the PC domain 
\cite{vosk}.

To elucidate this phenomenon, we employ the Landau relation
for the sp spectrum $\epsilon(p)$ in the specific form
\beq
{\partial \epsilon(p)\over\partial {\bf p}}=
{\partial \epsilon_0(p)\over\partial{\bf p}}
- {1\over 2}\int {\cal V}^{\pi}
({\bf p}-{\bf p}_1,\omega=0)
{\partial n({\bf p}_1)\over \partial {\bf p}_1} d\tau_1 \  ,
\label{lp}
\eeq
where $\epsilon_0(p)=p^2/2M^*_0$ is the regular part of the neutron spectrum
with $M^*_0\simeq 0.7M$, the customary value of the neutron effective
mass in the absence of spin-isospin fluctuations.  Upon straightforward
momentum integration, relation (\ref{lp}) yields
a closed RPA-like equation
\beq
\epsilon(p)=\epsilon_0(p) -
{1\over 2m_{\pi}^2}
\int\!\!\!\!\!\!\int\limits_{\,\,\,|p-p_1|}^{\,\,\,p+p_1}
\!\!\!\!\lambda^2_n(q){q^3\over p}\,D(q,0)\,
n(p_1)\,{p_1dp_1\,dq \over (2\pi)^2} \, ,
\label{ds2}
\eeq
well suited to investigation of the rearrangement of the Landau
state as the density climbs to the critical value $\rho_c$.  Results
of numerical calculations based on Eq.~(\ref{ds2}), depicted in Fig.~2,
indicate that the neutron Fermi surface becomes doubly connected at
$\eta_r=(\rho_c-\rho_r)/\rho_c\simeq 0.065$.

\begin{figure}[t]
\includegraphics[width=\linewidth]{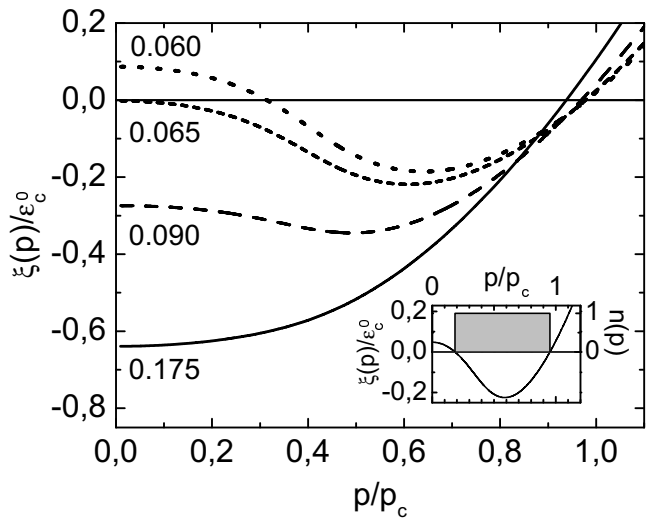}
\caption{
Neutron quasiparticle spectrum $\xi(p)$ in units of $\epsilon^0_c
=p_c^2/2M
=(3\pi^2\rho_c)^{1/3}/2M$,
plotted for different densities. Numbers near the
curves give the corresponding values of $\eta=(\rho_c-\rho)/\rho_c$.
Inset shows the quasiparticle spectrum $\xi(p)$ evaluated for
$\eta=0.062$, together with the associated quasiparticle momentum
distribution $n(p)$.
}
\label{fig:spec}
\end{figure}

With the redistribution of quasiparticles in momentum space, an 
inner neutron Fermi surface emerges at a low momentum $p_i$, enabling
momentum/energy conservation in the DU reactions and unleashing
rapid cooling.  When the neutron liquid becomes superfluid
under decreasing temperature, a triplet pairing gap $\Delta_i$ forms 
on the inner Fermi surface and applies the brakes to DU emission.
Even so, we find that this novel opportunity for DU cooling \cite{vosk} 
is not destroyed.  To gauge the neutron gap value $\Delta_i$ at 
the inner Fermi surface, we again employ Eq.~(\ref{bcs01}). 
Numerical computation yields 
\beq
\Delta_i\simeq \,\Delta(p_F)\,p_i/p_F \  .
\label{rel}
\eeq
The DU process may operate with full force only if the gap value 
$\Delta_n$ is markedly less than 100 keV.  According to Eq.~(\ref{rel}), 
this constraint is met at the inner Fermi surface if 
$p_i(\rho)\leq 0.02p_F$. The numerical calculation underlying 
Fig.~\ref{fig:spec} indicates that this inequality does 
hold in a thin shell of stellar material where 
$\rho-\rho_r\sim 5\times 10^{-4}\rho_c$.

Now consider, in outline, how our model explains experimental cooling 
data.  We assume that neutron stars such as RX J0822$-$4300 and
PSR B1055$-$52 have relatively low masses, such that pion condensation 
is not triggered in their cores.  We then attribute the high values of 
their surface temperatures to the absence of cooling mechanisms 
other than neutrino bremsstrahlung from $pp$ collisions, emission 
processes involving neutrons being switched off by substantial neutron 
$P$-pairing.

In explanation of the enhanced cooling rates of the Vela, Geminga, 
and 3C58 pulsars, we suggest that their central densities exceed 
some 0.5 fm$^{-3}$, and hence that a pion condensate occupies a 
significant portion of their interiors.  Rearrangement of the 
neutron quasiparticle distribution in a small region adjacent to 
the boundary between liquid and solid phases lifts the ban on the 
DU process.  However, the DU cooling rate, proportional to the volume
where this mechanism operates vigorously, is suppressed due to the 
restriction of the process to a narrow shell.  Consequently,
the corresponding $T_s$ values need not lie far below those of
the first group of neutron stars.  One feature of our model that 
warrants further examination is the fast transition from slow to 
enhanced cooling under increasing stellar mass.

Restriction of DU reactions to a narrow shell does not apply for the 
most massive neutron stars, in which the internal pressure becomes high 
enough to melt the pion-condensate lattice.  In the dense liquid core, 
the DU reactions are allowed to proceed apace even if the standard microscopic 
equation of state \cite{wir} is employed.  As yet, no observational 
evidence exists for such an unrestrained DU cooling mechanism. 

In this letter, we have proposed a new picture of the interior
of neutron stars based on the similarity between pion condensation
and the metal-insulator phase transition in the 2D strongly-correlated 
Fermi systems.  We have demonstrated that the incorporation of 
spin-isospin fluctuations has dramatic effects on the neutron and 
proton gaps in the liquid part of the stellar interior and induces 
a rearrangement of the neutron Fermi surface that triggers the 
direct-Urca reaction.  The associated cooling scenario provides for 
slow and accelerated cooling tracks that do not conflict with 
observational data. 

This research was supported in part by the National Science Foundation
under Grant No.~PHY-0140316, by the McDonnell Center for the Space
Sciences, and by Grant NS-1885.2003.2 from the Russian Ministry of 
Education and Science.  We thank G.~E.~Volovik and V.~M.~Yakovenko for
discussions, and D.~G.~Yakovlev for a critical reading of the manuscript.

\end{document}